\documentclass{aastex}
\usepackage{emulateapj5}
\usepackage{float,epsfig}
\usepackage{pstricks}

\newcommand{\mincir}{\raise
  -2.truept\hbox{\rlap{\hbox{$\sim$}}\raise5.truept \hbox{$<$}\ }}
\newcommand{\magcir}{\raise
  -2.truept\hbox{\rlap{\hbox{$\sim$}}\raise5.truept \hbox{$>$}\ }}

\begin{document}

\submitted{subm. to ApJ 2005, April 11}
\title{A Puzzling Merger in A3266: the Hydrodynamic Picture from XMM-Newton}
\author{A. Finoguenov$^{1,2}$, M. J. Henriksen$^2$,
F. Miniati$^3$, U. G. Briel$^1$, C. Jones$^4$}
\affil{
{$^1$ Max-Planck-Institut f\"ur extraterrestrische Physik,
             Giessenbachstra\ss e, 85748 Garching, Germany}\\
{$^2$ Joint Center for Astrophysics, Physics Department, University of
  Maryland, Baltimore County, Baltimore, MD 21250, USA}\\
{$^3$ Physics Department, ETH Z\"urich, CH-8093 Z\"urich, Switzerland} \\
{$^4$ Smithsonian Astrophysical Observatory, 60 Garden st., MS 2, Cambridge,
  MA 02138, USA}}

\begin{abstract}
Using the mosaic of nine XMM-Newton observations, we study the hydrodynamic
state of the merging cluster of galaxies Abell 3266. The high quality of the
spectroscopic data and large field of view of XMM-Netwon allow us to
determine the thermodynamic conditions of the intracluster medium on scales
of order of 50 kpc. A high quality entropy map reveals the presence of an
extended region of low entropy gas, running from the primary cluster core
toward the northeast along the nominal merger axis.  The mass of the low
entropy gas amounts to approximately $2\times 10^{13}$ solar masses, which
is comparable to the baryonic mass of the core of a rich cluster.  We test
the possibility that the origin of the observed low entropy gas is either
related to the disruption a preexisting cooling core in Abell 3266 or to the
stripping of gas from an infalling subcluster companion.  We find that both
the radial pressure and entropy profiles as well as the iron abundance of
Abell 3266 do not resemble those in other known cooling core clusters (Abell
478). Thus we conclude that the low entropy region is subcluster gas in the
process of being stripped off from its dark matter halo. In this scenario
the subcluster would be falling onto the core of A3266 from the foreground.
This would also help interpret the observed high velocity dispersion of the
galaxies in the cluster center, provided that the mass of the subcluster is
at most a tenth of the mass of the main cluster.
\end{abstract}


\section{Introduction}

Current theories of structure formation are based on hierarchical models
whereby small structures form first and, by assembly, build up the larger
structures. In this picture, mergers play an important role, not only in
driving the formation of galaxy clusters, but also in affecting the
properties of the intra-cluster medium such as their thermodynamic
conditions, metal content and the like.  The occurrence of a merger event
results in dramatic consequences for the systems that take part in it. As a
large amount of energy is released in the process, the intracluster medium
is strongly stirred up which results in the production of shocks and
turbulence (Miniati et al. 2000, Schuecker et al. 2004).  Furthermore, the
trans/super-sonic infalling subsystem may be stripped off of their gas by
ram pressure work (Gunn \& Gott 1972).  This mechanism may supply the main
cluster with low-entropy gas, and perhaps with high metalicity too (Motl
et al. 2004).  In addition, disruption of cool condensations and abundance
gradients may result in the cluster center. Until now, however, the evidence
the latter has been only {\it ex post facto} in that post-merger systems
appear to lack central cool gas reservoirs and central abundance
enhancements so that little is known of the intermediate evolutionary
stages.

In this paper we study the properties of the intracluster medium in Abell
3266 (also known as Sersic 40-6), a well studied cluster merger with
redshift of 0.0594, but still not fully understood.  A statistical analysis
of the galaxy kinematical data (Quintana et al. 1996) show evidence of a
merger.  They show that the velocity dispersion in the center, $\sim$1300 km
s$^{-1}$, is significantly higher than the global value of 1000 km s$^{-1}$.
This was interpreted as evidence of a relatively localized merger in
progress that has increased the internal energy of the cluster primarily in
the central region.  However, no more than 10\% of the galaxies appear to be
falling into the cluster along the line-of-sight and this is at a velocity
of less than 1000 km s$^{-1}$; much less than typical velocities associated
with low Mach numbers shocks in the cluster center.

The central cD galaxy in Abell 3266 is in a dumbbell morphology. Multiple
nuclei are found in over 25\% of the first rank galaxies in rich clusters
(Hoessel 1980). The relative velocity of the dumbbell, $400\pm39$ km$^{-1}$,
is larger than the stellar velocity dispersion of the cD nucleus, $327\pm
34$ km$^{-1}$, and may indicate that the pair was formed in the
merger. However, the asymmetric rise in the stellar velocity dispersion of
the cD peaks at around 700 km$^{-1}$ (Carter et al. 1985) suggesting that
the cD is tidally disturbed by a very massive object, more massive than the
second nucleus.

Evidence of a merger in A3266 is also supported by X-ray data.  This is
provided by both the temperature structure of the ICM as well as the
elongated morphology of the surface brightness of the system, as seen by
ASCA (Henriksen, Donnelly, \& Davis 2000) .  These findings have recently
been confirmed by {\it Chandra} data (Henriksen \& Tittley 2002).  In
addition, the {\it Chandra} hardness map also suggests a cool, low entropy
region running along the surface brightness elongation (the merger axis).

The temperature structure around the cluster center observed by {\it
Chandra} could be produced by the propagation of a shock induced by the
passage, off-center and in the plane of the sky, of an infalling clump
during the initial phases of a merger (Henriksen \& Tittley 2002). However,
it is not clear yet whether this is the actual scenario because the geometry
of the merger cannot be established conclusively by the current optical
data.  In any case, the extent and location of the low entropy region could
hint at a link between the cD galaxy and the merger.  The plume of cooler
gas could be stripped material from the cD galaxy, or a disrupted cooling
flow centered on the galaxies (Henriksen \& Tittley 2002).

We have carried out analysis of XMM-Newton data of A3266.  Given high
quality of the data and the employment of newly developed reduction
techniques (Briel et al. 2004), we are able to produce accurate maps that
describe the thermodynamic state of the ICM in A3266.  While the qualitative
picture is confirmed, the new data provides a much more detailed picture of
the merging process.  This may serve as testbed to our understanding of the
role of gas stripping on cluster scales.

The paper is organized as follows, in \S2 we describe the observation, in
\S3 we present the results, followed by a discussion in \S4.

We adopted a LCDM cosmology with $\Omega_{m}=0.3$, $\Omega_{\Lambda}=0.7$
and $H_{\rm o}=70$~km~s$^{-1}$~Mpc$^{-1}$, which leads to a luminosity
distance of $D_L=210$ Mpc and an angular scale of 56 kpc/arcminute for
A3622.

\section{Data}

XMM-Newton (Jansen et al. 2001) has observed A3266 as a part of the
GTO program of instrumental scientists at MPE. Tab.\ref{t:ol} details
the mosaic of the cluster, where column (1) is the name of the
proposed field, (2) is the assigned XMM archival name, (3) R.A. and
Decl. of the pointing, (4) net Epic-pn exposure after removal of
flaring episodes, (5) XMM-Newton revolution
number. All Epic-pn observations were performed using the extended
full frame mode with the frame integration time of 199 ms and used
Medium filter.

%\begin{table}[H]
%\begin{minipage}{8cm}
%\footnotesize
\begin{center}
\renewcommand{\arraystretch}{1.1}\renewcommand{\tabcolsep}{0.12cm}
\tabcaption{\footnotesize
\centerline{XMM Epic-pn log of A3266 cluster observation.}
\label{t:ol}}

\begin{tabular}{lccccc}
\hline
\hline
      &  Obs.       & Pointing  & net       & XMM \\
      &   ID        & R.A Decl. & exp.      & Orbit\\
      &             & (Eq.2000) & ksec      & \\
\hline
f1 &0105260701 & 68.11125 -61.31277 & 13.6 & 149\\
f2 &0105260801 & 67.81583 -61.25916 & 15.2 & 154\\
f3 &0105260901 & 67.59833 -61.50888 & 17.8 & 153\\
f4 &0105261001 & 67.55541 -61.35055 &  4.3 & 147\\
f5 &0105261101 & 68.14791 -61.48361 &  6.6 & 146\\
f6 &0105262001 & 67.90416 -61.56722 &  1.9 & 145\\
f5 &0105262101 & 68.14791 -61.48361 &  3.5 & 146\\
f4 &0105262201 & 67.55541 -61.35055 &  2.9 & 147\\
f6 &0105262501 & 67.90416 -61.56722 &  3.5 & 598\\
\hline
\end{tabular}
\end{center}
%\end{minipage}
%\end{table}

The initial steps of data reduction are similar to the procedure tested on
other XMM-Newton mosaics and is described in detail in Briel, Finoguenov,
Henry (2004). Throughout our analysis we used XMMSAS 6.1. Details of our
light curve screening could be found in Zhang et al. (2004). The analysis
consists of two parts: estimating the image and temperature structure of the
cluster and verifying it through the spectral analysis. The first part
consists in producing temperature estimates, based on the calibrated
wavelet-prefiltered hardness ratio maps and producing the projected pressure
and entropy maps. Wavelet filtering is used to find the structure and
control its significance. The background is considered differently in
imaging and spectral analysis. In the first we use the in-field estimate of
the background for every pointing, using events furthest from the optical
axis of the telescope and bright emission zones of the A3266, where
instrumental background should dominate and then subtract this background
using the exposure maps with no vignetting normalized to reproduce the
background level in the region selected for the background estimate.

The second, spectroscopy part of the analysis, uses the mask file, created
based on the results of hardness ratio analysis described above. First
application of this technique are presented in Finoguenov et al. (2004).
For background removal we use the background accumulation, obtained with the
medium filter, while we first refined our background subtraction by modeling
the residual background observed at outskirts of A3266. We use the 0.5--7.9
keV band in the spectral analysis.

In our spectral analysis we use the APEC plasma code (Smith et al. 2001) to
measure the temperature, element abundance (assuming the solar abundance
ratio of Anders \& Grevesse 1989) and emission measure. By making an
estimate of the volume occupied by each emission zone it is then possible to
recover the pressure and the entropy in absolute units, as detailed in Henry
et al. (2004).

\section{Results}

Advantage of XMM-Newton over the previous missions consists in the ability
to provide detailed temperature maps, which allows to study the pressure and
entropy state of the cluster gas. In Finoguenov et al. (2004) and Briel et
al. (2004) we have developed a simple technique for selecting the regions
for the spectral analysis, using both image and hardness ratio as an
input. The XMM exposure of A3266 allows to produce almost a hundred of
spatially independent temperature estimates.

\begin{figure*}
\includegraphics[width=8.cm]{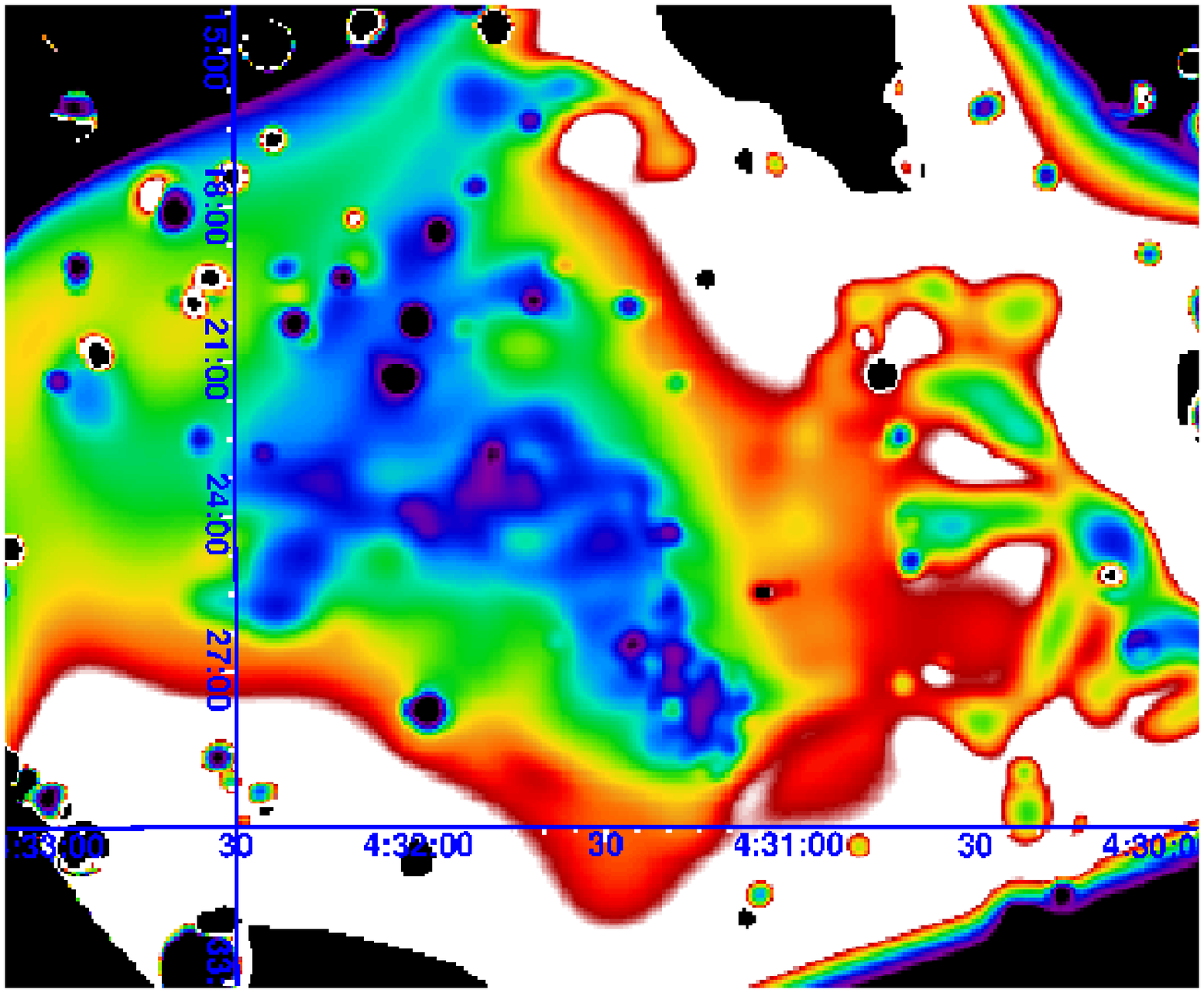}\hfill\includegraphics[width=8.cm]{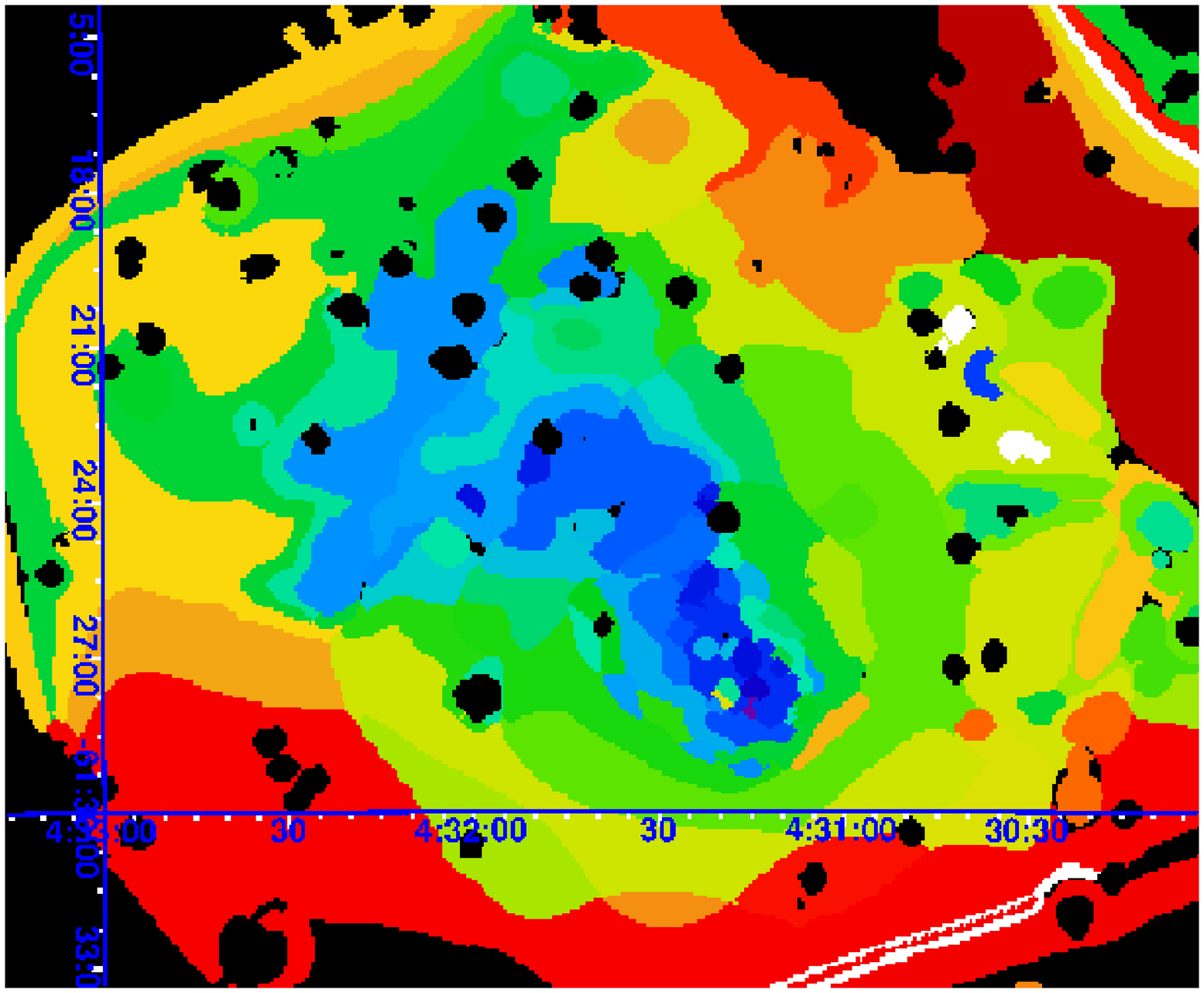}\\
\includegraphics[width=16cm]{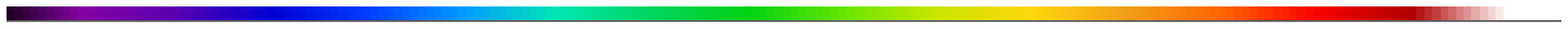}

\figcaption{{\it Left panel}. Wavelet ({\it left }) and spectroscopic ({\it
    right }) reconstructions of the projected entropy.
\label{f:ent}}
\includegraphics[width=8.cm]{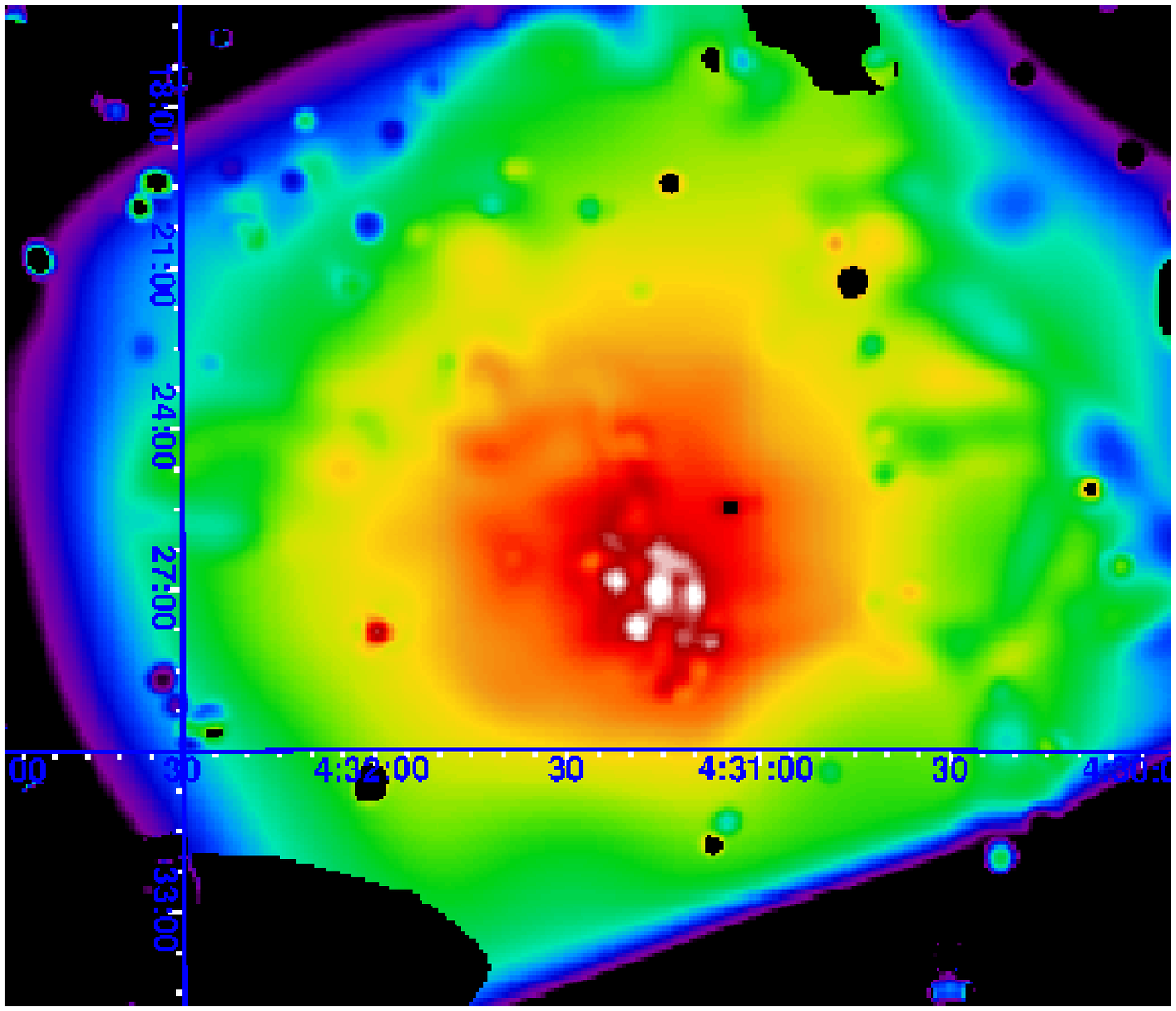}\hfill\includegraphics[width=8.cm]{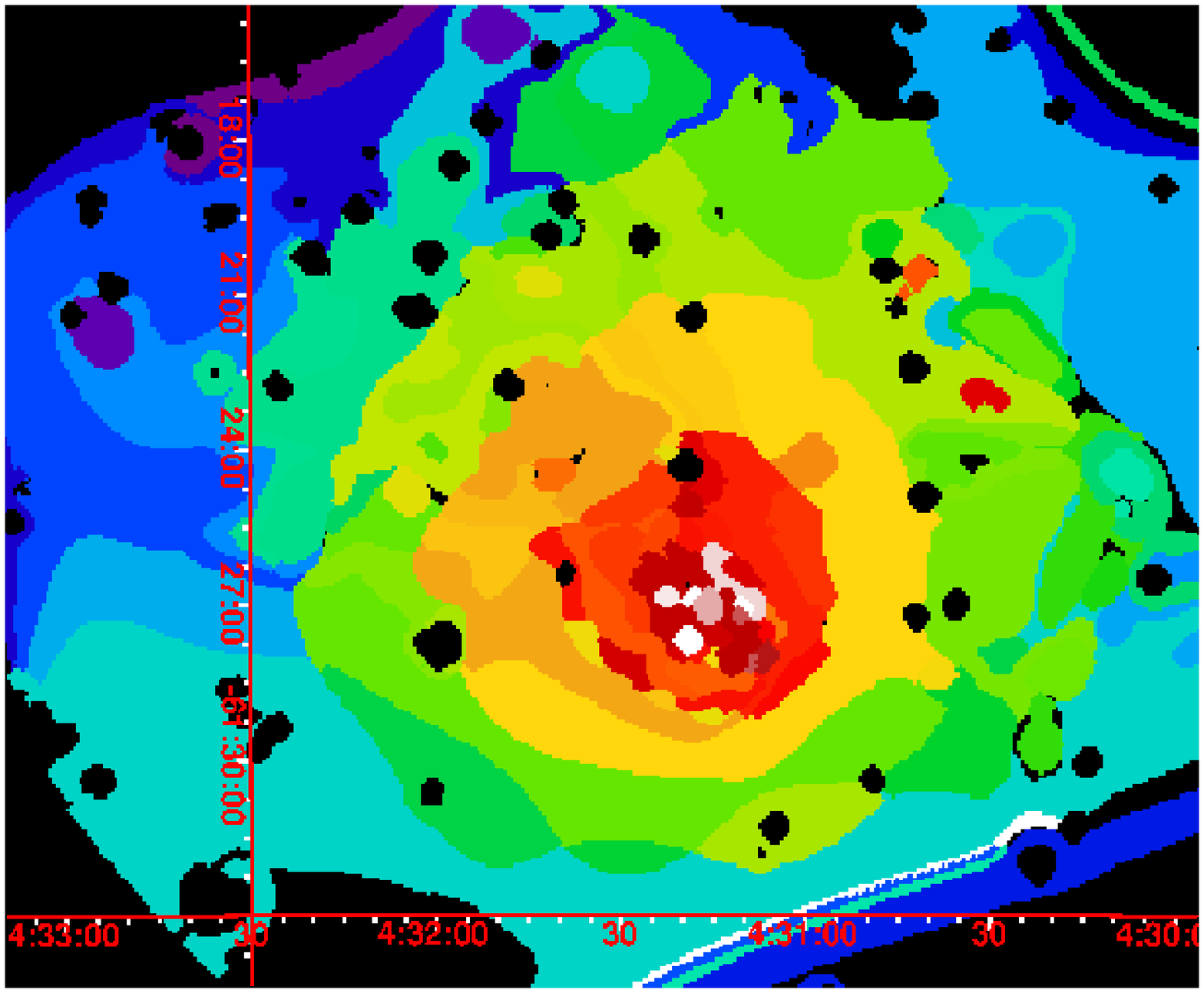}\\
\includegraphics[width=16cm]{plots/cbar_sls.ps}

\figcaption{{\it Left panel}. Wavelet ({\it left }) and spectroscopic ({\it
    right }) reconstructions of the  the projected pressure.
\label{f:nkt}}
\includegraphics[width=8.cm]{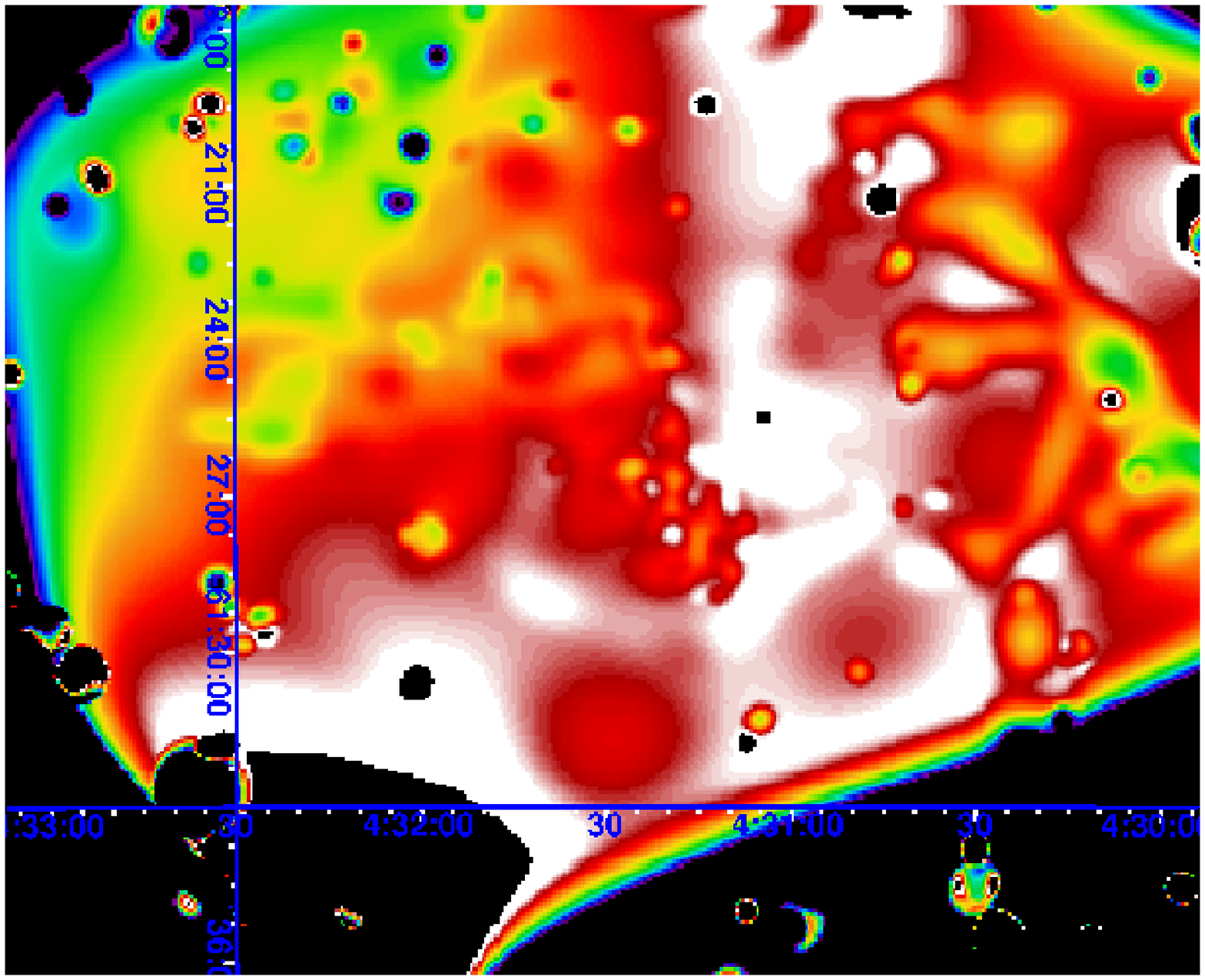}\hfill\includegraphics[width=8.cm]{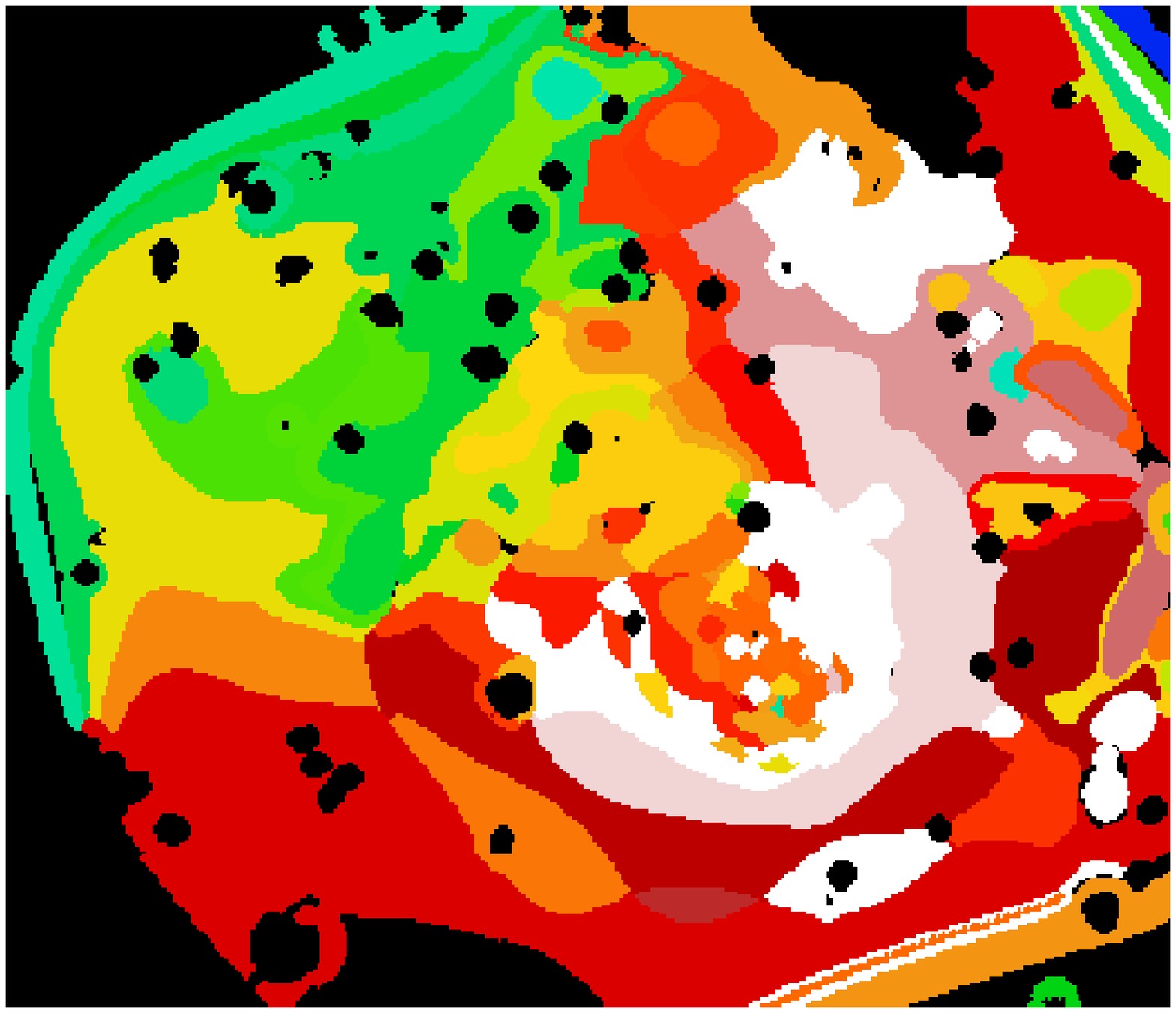}\\
\includegraphics[width=16cm]{plots/cbar_sls.ps}

\figcaption{{\it Left panel}. Wavelet ({\it left }) and spectroscopic ({\it
    right }) reconstructions of the spatial temperature distribution. The
  color legend is as follows: green, yellow, red, white
  corresponds to the temperature of 4, 6, 7, 9 keV, respectively.
\label{f:te}}
\end{figure*}

Fig.\ref{f:ent}--\ref{f:te} show maps of entropy, pressure and temperature,
respectively, as obtained from the wavelet analysis (left) as well as from
direct spectral analysis (right).  The entropy map reveals the presence of a
giant plume of low entropy gas (LEG hereafter) extending over $15^\prime$ to
the north-east from the center.  This is interesting because the distinct
value of the entropy of the LEG suggests a common origin for it.  The
orientation of the LEG is aligned with the major axis of the merger,
i.e. the axis of the elongated X-ray morphology. This strongly suggests an
association for the origin of the LEG with the merger itself.  The question
is whether the LEG belongs to a cool core in A3266, now undergoing
disruption, or it is just gas of an infalling subcluster undergoing ram
pressure stripping. This will be addressed in the discussion section.  The
entropy in the LEG has significant structure on small but well resolved
scales.  In particular, local enhancements are clearly visible: these are
either pre-existed as contact discontinuities due to inefficient mixing or
have been recently produced by relatively weak shocks. In fact, all
twelve major entropy enhancements have a corresponding pressure
enhancement. Finally, these maps show a great number of the entropy dips in
both the tail and overall: they are aligned with galaxies and galaxy
subgroups, revealing a highly incomplete mixing of the gas in the merger. A
detailed comparison of the entropy and dynamics of the cluster will be
presented elsewhere.

%The pressure map, on the other hand, shows, grossly speaking, a rather
%isotropic distribution. 
The core region ($r<0.2r_{500}$) exhibits fluctuations at the level of
$49\pm7$\% for the entropy and $55\pm2$\% for the pressure, in respect to a
power law approximation of the radial profile. In the central most
$0.06r_{500}$, fluctuations of entropy are similar ($51\pm7$\%), while
fluctuations of the pressure are higher ($105\pm3$\%). Signatures of
disturbances are also visible in the outer regions ($0.2<r/r_{500}<0.6$),
particularly towards the NE. The R.M.S. of fluctuations amounts to
$48\pm8$\% for the entropy and $44\pm2$\% for the pressure and are dominated
by the asymmetries in the azimuthal distributions.

Finally, we comment on the temperature map, even though it is not
independent of the previous two quantities. Fig. 3 reveals a slight
temperature enhancement ahead of the region covered by the LEG, which may be
due to gas compression and occasionally weak shocks as well. The structure
seen in the entropy map is also reflected in the temperature. In the core
region, temperature enhancements are also clearly seen in the pressure map,
suggesting strong interactions leading to the formation of small scale
shocks.

The results presented in the above maps are summarized and better quantified
in Tab.\ref{t:spe}. There, in analogy to the procedure followed for the
temperature map, we divide the observed region of A3266 in 33 regions with
the same spectral temperature to within 1 keV.  Fig.\ref{f:lab} shows such
regions. For each one of them, in Tab.\ref{t:spe} we list the measured
physical properties with their $\pm1\sigma$ errors. In particular, Col. (1)
and (2) corresponds to the zone labels, (3) to the temperature in keV, (4)
iron abundance as a fraction of solar photospheric value of Anders \&
Grevesse (1989). Derived quantities, that use an estimation of the projected
length, as described below are reported in cols.(5--7). These are entropy,
pressure and the gas mass. No account for the gas mass not associated with
the directly observed component was attempted. The numbers given in the
table confirm that LEG consists of lower entropy and somewhat higher Fe
abundance gas. In the very back of the LEG (regions 22, 30, 33) this
separation becomes marginal.

\includegraphics[width=8.cm]{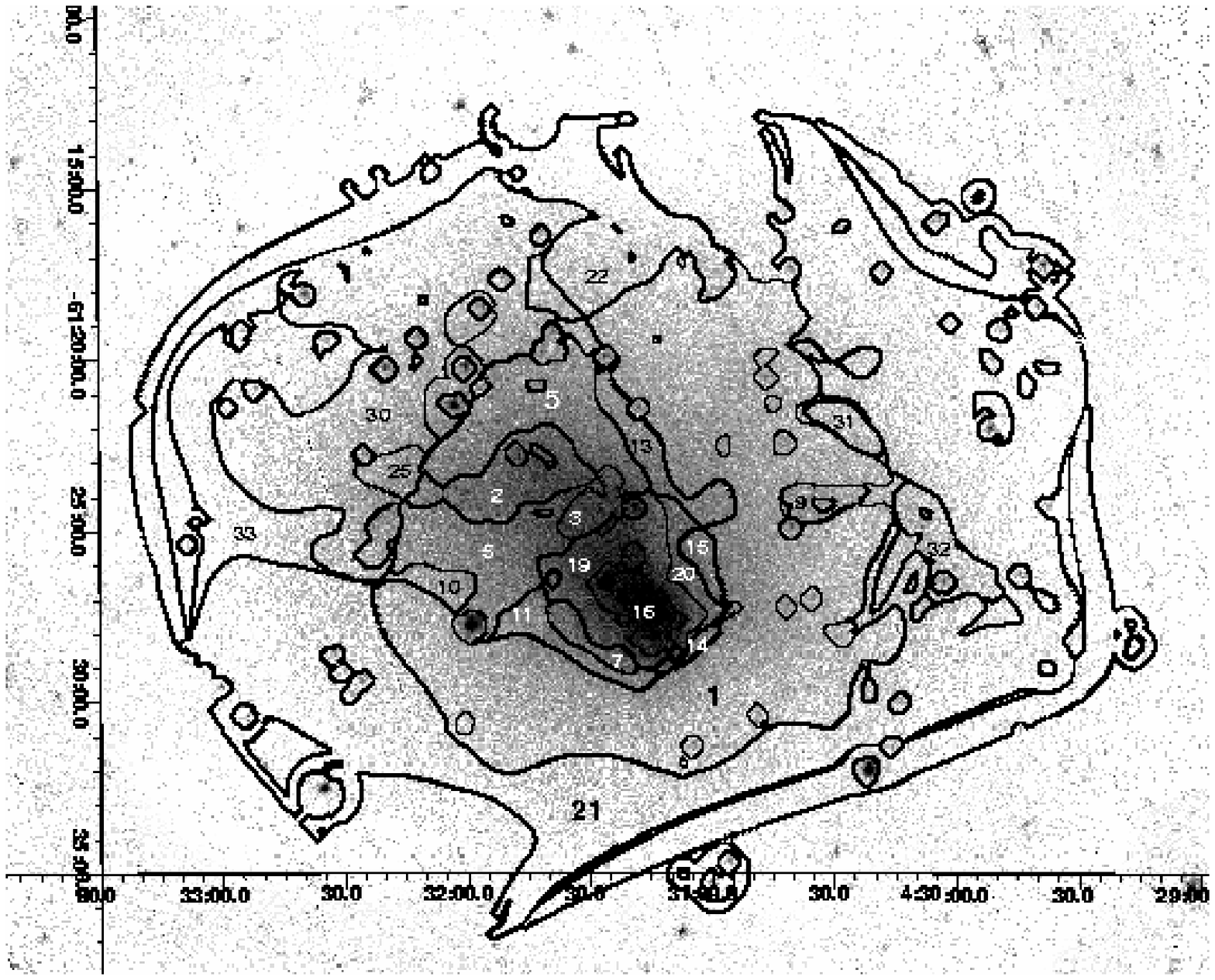}
\figcaption{Background subtracted, exposure and vignetting corrected image
  of A3266 in the 0.5--2 keV band. Contours indicate the zones of spectral
  extraction.
\label{f:lab}}

%\begin{table*}[ht]
{
\begin{center}
\footnotesize
{\renewcommand{\arraystretch}{0.9}\renewcommand{\tabcolsep}{0.05cm}
\tabcaption{\footnotesize
Properties of main regions of A3266.
\label{t:spe}}

\begin{tabular}{cccccccc}
 \hline
 \hline
N & Name &$kT$ & $Fe/Fe_\odot$ & S & P, $10^{-11}$ & $M_{\rm gas}$\\
  &  & keV     &              & keV cm$^2$ & erg cm$^{-3}$ & $10^{11}
M_\odot$  \\
 \hline
\multicolumn{5}{c}{low entropy gas (LEG)}\\
16 &head& $5.1\pm0.1$&$0.22\pm0.03$&$157\pm3$&$4.7\pm0.1$&4.7 \\
19 &core& $5.8\pm0.1$&$0.18\pm0.02$&$253\pm5$&$3.2\pm0.07$&13.6 \\
 3 &striped core-1& $6.0\pm0.2$&$0.30\pm0.05$&$264\pm10$&$3.3\pm0.1$&2.7 \\
 2 &striped core-2& $4.3\pm0.1$&$0.20\pm0.03$&$269\pm7$&$1.4\pm0.1$&12.0 \\
25 &striped core-3& $4.4\pm0.3$&$0.24\pm0.10$&$324\pm26$&$1.1\pm0.1$&2.0 \\
 7 &SW core side& $6.1\pm0.4$&$0.19\pm0.07$&$279\pm16$&$3.1\pm0.2$&1.8 \\
11 &S-edge of core& $6.2\pm0.3$&$0.11\pm0.06$&$375\pm20$&$2.1\pm0.1$&3.8 \\
20 &shock-1& $7.5\pm0.4$&$0.29\pm0.06$&$355\pm18$&$3.7\pm0.2$&3.1 \\
15 &shock-2& $8.2\pm0.6$&$0.45\pm0.13$&$367\pm29$&$4.4\pm0.4$&1.2 \\
14 &shock-3& $9.2\pm1.5$&$0.10\pm0.20$&$372\pm64$&$5.7\pm1.0$&0.4 \\
 5 &tail-1& $4.9\pm0.1$&$0.21\pm0.02$&$351\pm5$&$1.3\pm0.02$&38.6 \\
30 &tail-2& $4.1\pm0.1$&$0.24\pm0.03$&$528\pm10$&$0.44\pm0.01$&64.0 \\
33 &tail-3& $4.3\pm0.2$&$0.22\pm0.07$&$653\pm40$&$0.38\pm0.02$&18.2 \\
10 &S-edge of tail& $6.9\pm0.8$&$0.06\pm0.10$&$527\pm59$&$1.6\pm0.2$&2.2 \\
13 &N-edge& $5.5\pm0.3$&$0.14\pm0.05$&$383\pm18$&$1.5\pm0.1$&6.3 \\
22 &N-edge-2& $5.0\pm0.2$&$0.26\pm0.06$&$709\pm33$&$0.48\pm0.02$&17.3 \\
\multicolumn{5}{c}{other zones}\\
 1 &main& $5.8\pm0.1$&$0.13\pm0.01$&$556\pm8$&$1.0\pm0.01$&124. \\
21 &outskirts& $4.7\pm0.1$&$0.14\pm0.03$&$701\pm16$&$0.42\pm0.01$&129. \\
 9 &W group& $4.8\pm0.3$&$0.15\pm0.10$&$351\pm26$&$1.2\pm0.1$&2.0 \\
31 &W group-2& $5.9\pm0.6$&$0.28\pm0.14$&$457\pm48$&$1.4\pm0.2$&1.5 \\
32 &W group-3& $5.3\pm0.3$&$0.29\pm0.11$&$503\pm32$&$0.90\pm0.1$&4.5 \\
\hline
\end{tabular}
}
\end{center}
}
%\end{table*}

\section{Discussion and conclusions}

\begin{figure*}
\includegraphics[width=16.cm]{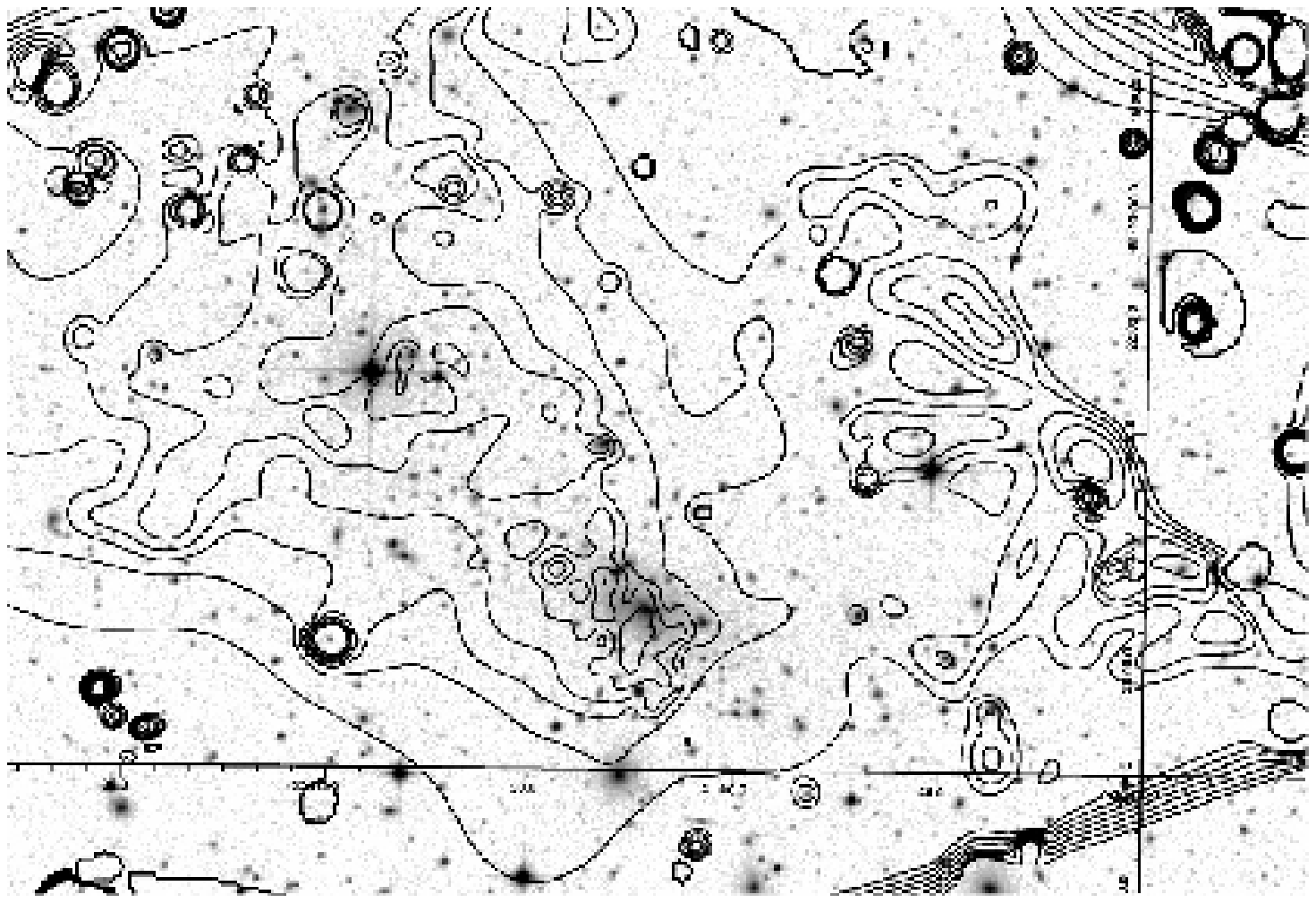}
\figcaption{DSS-2 optical image (red) of A3266. Coordinates are shown for
  Equinox J2000. Contours indicate the levels of equal projected entropy
  starting at the minimum level observed at the core with contours
  increasing by a factor of 1.2.
\label{f:dss}}
\end{figure*}

The sum of the mass of the low entropy regions near the cluster core termed
LEG in Table 2 indicates a total mass of $2\times10^{13} M_\odot$.  This
mass is comparable to the core gas mass (r $<$ 250 kpc) of clusters with
mass similar to Abell 3266 (Peres et al. 1998). It is also comparable to the
total gas masses of cool, low mass clusters (Mohr, Mathiesen, \& Evrard
1999), or to high mass groups (Mulchaey et al. 1996). Thus, the extended
feature of low entropy gas is either the core of the main cluster or gas
stripped off of a secondary merger component with a secondary-to-primary
mass ratio of 1:10.

To tie the entropy map to the dynamics of the cluster we have produced in
Fig. \ref{f:dss} an overlay of the DSS-2 red (optical) image of the A3266
with contours of equal entropy.  Interestingly, the central concentration of
galaxies, including the cD, lies well within the low entropy region. In
particular the head of the LEG region is located at the position of the cD,
although slightly shifted towards south-east.  Therefore, based on their
corresponding position on the sky, one would be led to straightforwardly
associate the cool, low entropy gas with the cD and its surrounding
galaxies.  Thus, either the cool gas belongs to A3266 in which case it may
suggest the pre-existence of a cool core there or, alternatively, the cD
galaxy is a part of the infalling subclump. The intracluster gas from the
lower mass secondary subcluster is typically characterized by the lower
temperature.

\begin{figure*}
\includegraphics[width=8.cm]{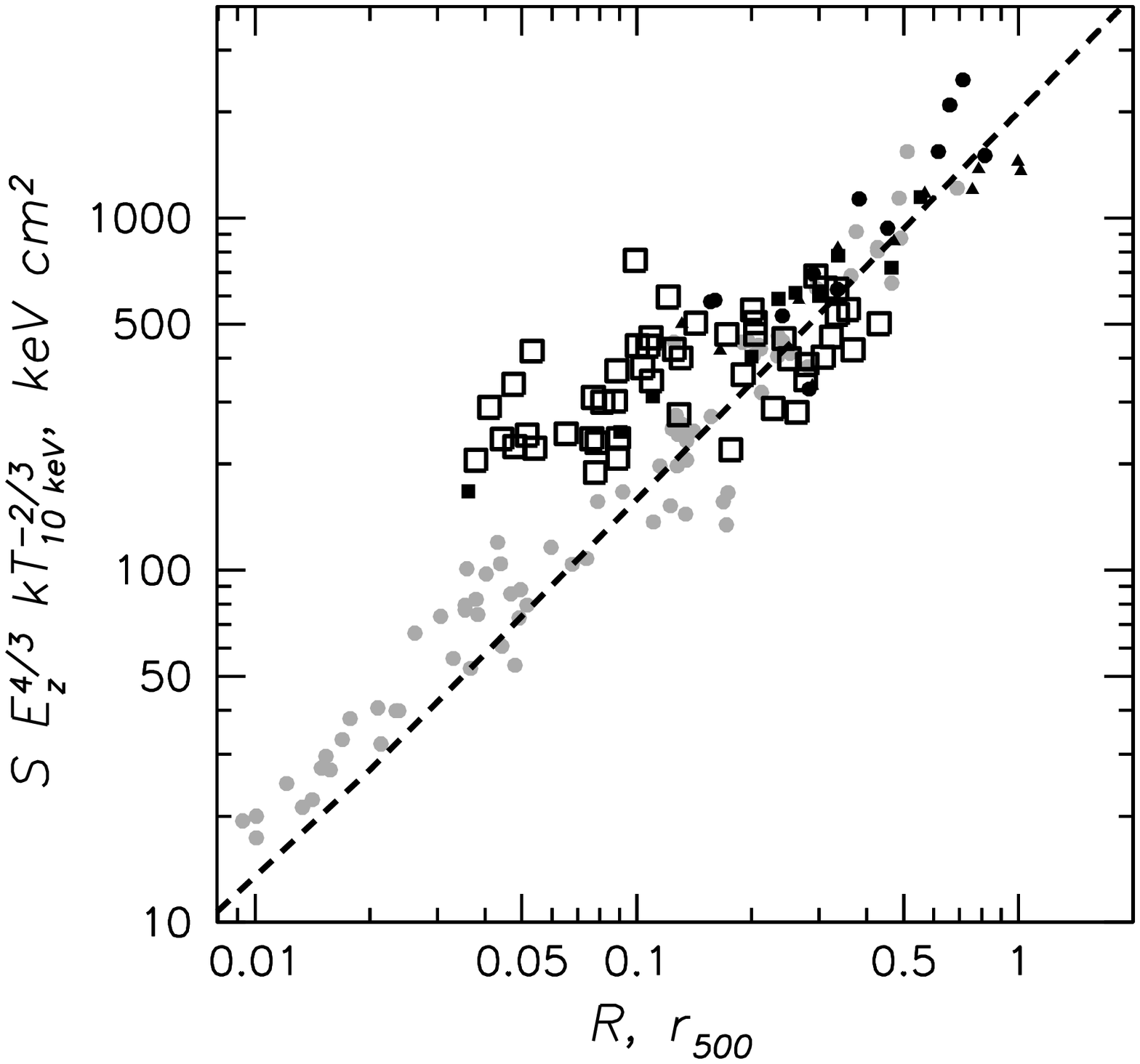}\hfill
\includegraphics[width=8.cm]{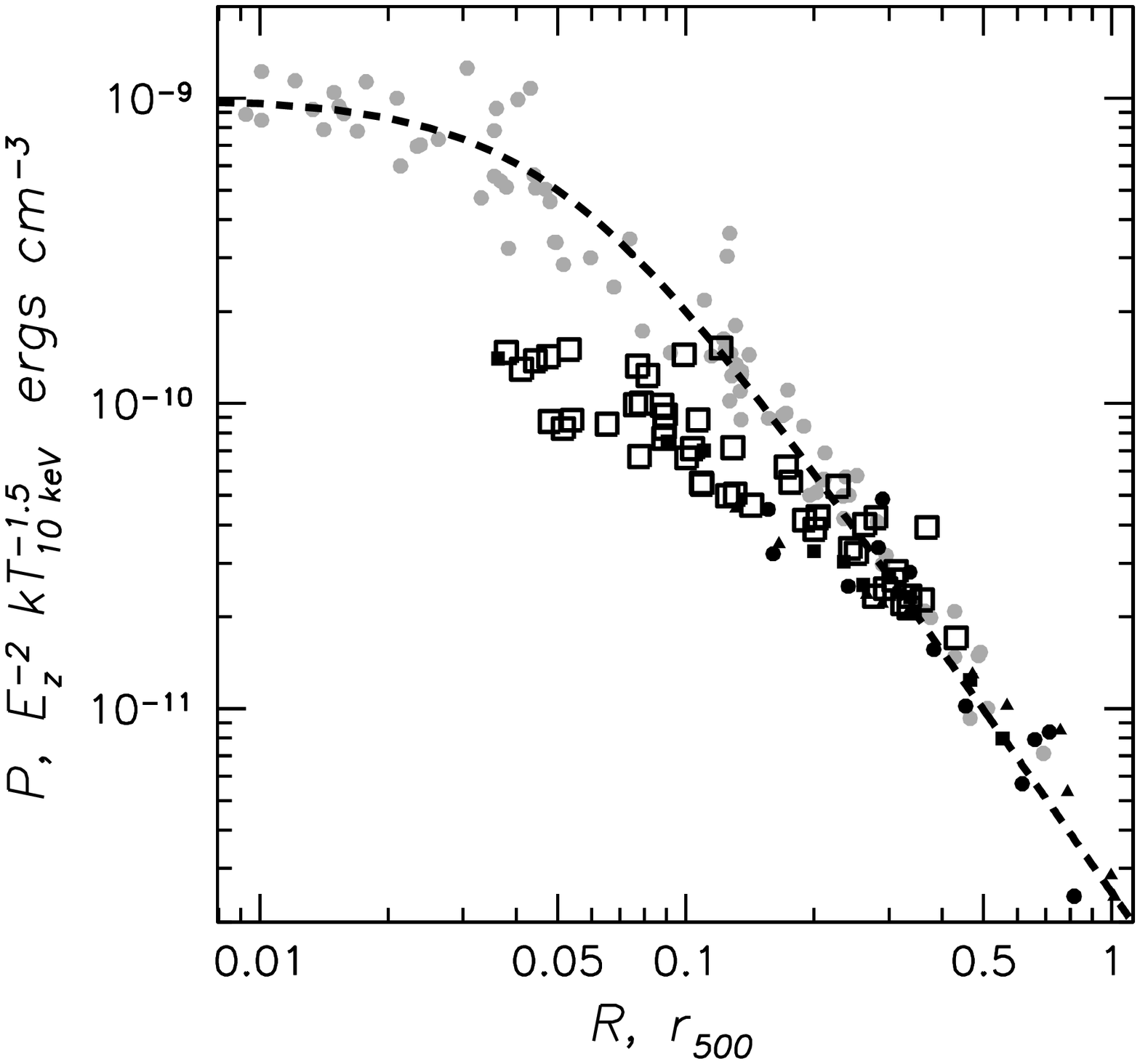}
\figcaption{Comparison of
the entropy ({\it left panel}) and pressure ({\it right panel}) profiles of
thermal electrons in A3266 (open squares) with examples of cooling flow
cluster (A478, grey filled circles) and merging clusters (black filled
symbols) A754 (triangles), A3562 (squares), A3667 (circles). Dashed line
indicates the $r^{1.1}$ law expected from hierarchical buildup of clusters
(Tozzi \& Norman 2001; Voit 2004), normalized to the scaling results of
Ponman et al. (2003).
\label{f:pro}}
\end{figure*}

To test these hypothesis, in Fig.\ref{f:pro}, we plot the entropy and
pressure radial profile for Abell 3266 together with several comparison
clusters: A3562 in Shapley supercluster, A754, A3667 (Finoguenov et
al. 2005), which are merging clusters and Abell 478, which is a cooling core
cluster. As Fig. 5 shows, compared to Abell 478, the LEG region in A3266
exhibits a quite different radial profile in both entropy and pressure. At
small radii, the entropy in the LEG region is systematically higher and the
pressure is systematically lower. This suggests that, if the LEG originated
in the primary cluster, then it would not be simply a stripped, pre-existing
cool core.  Also, the iron abundance in the LEG region is similar to that in
the ICM of Coma, which does not have a cooling core (Arnaud et al. 2001),
and is also similar to the rest of the cluster. In contrast, De Grandi et
al. (2004) point out that central abundance enhancements are always found in
cool cores and are associated with central galaxies. The fact that the
abundance in the LEG is similar to that in the rest of the cluster further
argues against its origin in a cool core.

The entropy and pressure profiles of the LEG region are more similar to the
comparison clusters that are merging clusters than the cooling flow region
of Abell 478 and may therefore be composed largely of the intracluster
medium of the secondary cluster. This is generally consistent with
hydrodynamical simulations showing that intracluster gas from a low mass
secondary may be stripped as it passes through the high mass, high density
core.  Nonetheless, it is quite puzzling that the cD and the other core
galaxies sit on top of the leading edge of the LEG. It is conceivable that
we are seeing things in projections so that the LEG is falling onto the core
of A3266 from the foreground.  This scenario may not be as unlikely after
all, as the infalling subclump aims at the core of the companion cluster (so
as to produce a superposition along the line of sight). In addition, such a
configuration may ease some of the optical arguments. In particular, this
may explain the increasing apparent velocity dispersion of galaxies towards
the center (Quintana et al. 1996; Henriksen, Donnelly, Davis 2000), as being
caused by a displacement between the core of the main cluster and the
infalling cluster.  In order for this to be consistent with the optical data
the mass of the subcluster must be in the ratio of no more than 1:10 with
respect to the main cluster.

Alternatively, the morphology of the "leading edge" of the LEG (i.e.,
compressed isophotes spatially coincident with the cD galaxy) could
simply be due to sloshing of the core after an off-center merger
(Tittley \& Henriksen 2005).

Finally we note that Abell 3266 may have a more complex merger history than
the single 1:10 mass ratio merger that has been proposed. On the west side
of the cluster, there is an extended collection of entropy debris. Within
the debris are several peaks that appear to be associated with galaxies (see
Figure \ref{f:dss}). The masses of the peaks are comparable to group masses
(Mulchaey et al. 1996).  This second region of low entropy gas may indicate
an additional episode of merging as A3266 is the most prominent cluster in
the southern Horologium cluster concentration.

%Rethinking the optical arguments, a possibility of an alignment in
%the LOS of two merging systems arise. This reveals itself through an
%increasing apparent velocity dispersion of galaxies towards the
%center, caused by a displacement between the core of the main cluster
%and the infalling cluster. 

\section*{Acknowledgments}
This paper is based on observations obtained with XMM-Newton, an ESA
science mission with instruments and contributions directly funded by
ESA Member States and the USA (NASA). The XMM-Newton project is
supported by the Bundesministerium f\"{u}r Bildung und
Forschung/Deutsches Zentrum f\"{u}r Luft- und Raumfahrt (BMFT/DLR),
the Max-Planck Society and the Heidenhain-Stiftung, and also by PPARC,
CEA, CNES, and ASI.  AF and FM thank the Joint Astrophysical Center of
the UMBC for the hospitality during their visit. FM acknowledges support
from the Zwicky Prize Fellowship program at ETH-Z\"urich.

\end{document}